\documentclass[a4paper,10pt]{article}
\usepackage{graphicx}
\usepackage{comment}
\usepackage{amssymb}
\usepackage{amsmath}
\usepackage{nicefrac}
\usepackage{graphicx}
\usepackage{dcolumn}
\usepackage{bm}
\usepackage{comment}
\usepackage{multirow}
\usepackage{cite}
\usepackage{xcolor}
\usepackage{url}
\usepackage{mathrsfs}
\usepackage{supertabular}
\usepackage[normalem]{ulem}
\usepackage{lscape}
\usepackage{soul}
\usepackage{subcaption} 
\usepackage{bold-extra}

\newcommand{\qbin}[2]{\left[\begin{array}{c}\!#1\!\\\!#2\!\end{array}\right]_q}

\begin{document}
            
\newcommand{\bin}[2]{\left(\begin{array}{c}\!#1\!\\\!#2\!\end{array}\right)}
\newcommand{\threej}[6]{\left(\begin{array}{ccc}#1 & #2 & #3 \\ #4 & #5 & #6 \end{array}\right)}
\newcommand{\sixj}[6]{\left\{\begin{array}{ccc}#1 & #2 & #3 \\ #4 & #5 & #6 \end{array}\right\}}
\newcommand{\regge}[9]{\left[\begin{array}{ccc}#1 & #2 & #3 \\ #4 & #5 & #6 \\ #7 & #8 & #9 \end{array}\right]}
\newcommand{\La}[6]{\left[\begin{array}{ccc}#1 & #2 & #3 \\ #4 & #5 & #6 \end{array}\right]}
\newcommand{\hj}{\hat{J}}
\newcommand{\hux}{\hat{J}_{1x}}
\newcommand{\hdx}{\hat{J}_{2x}}
\newcommand{\huy}{\hat{J}_{1y}}
\newcommand{\hdy}{\hat{J}_{2y}}
\newcommand{\huz}{\hat{J}_{1z}}
\newcommand{\hdz}{\hat{J}_{2z}}
\newcommand{\hup}{\hat{J}_1^+}
\newcommand{\hum}{\hat{J}_1^-}
\newcommand{\hdp}{\hat{J}_2^+}
\newcommand{\hdm}{\hat{J}_2^-}

\huge

\begin{center}
Statistics of multi-electron states and $J$-levels in atomic configurations
\end{center}

\vspace{0.5cm}

\large

\begin{center}
Jean-Christophe Pain$^{a,b,}$\footnote{jean-christophe.pain@cea.fr} and Xavier Blanc$^{c}$
\end{center}

\normalsize

\begin{center}
\it $^a$CEA, DAM, DIF, F-91297 Arpajon, France\\
\it $^b$Universit\'e Paris-Saclay, CEA, Laboratoire Mati\`ere en Conditions Extr\^emes,\\
\it 91680 Bruy\`eres-le-Ch\^atel, France\\
$^c$Université Paris Cité, Sorbonne Université, CNRS, Laboratoire Jacques-Louis Lions, F-75013 Paris, France\\
\end{center}

\vspace{0.5cm}

\begin{abstract}
The number and nature of atomic configurations are cornerstones of atomic spectroscopy, especially for the calculation of hot-plasma radiative properties. The knowledge of the distributions of magnetic quantum number $M$ and angular momentum $J$ for $N$ identical fermions in a subshell with half-integer spin $j$ is a prerequisite to the determination of the structure of such configurations. The problem is rather complicated, since the possible occurrence of a specific values of $J$ is governed by the Pauli exclusion principle. Several methods, such as generating functions, recurrence relations or algebraic number theory —for instance via Gaussian polynomials— have proven effective in addressing this issue. However, up to now, no general formula was known. In the present work, we present exact and compact explicit formulas for the number of atomic configurations and for the distributions of the total magnetic quantum number $M$ and angular momentum $J$.
\end{abstract}

\section{Introduction}\label{sec1}

The knowledge of the number of atomic configurations (i.e., the number of possible ways to distribute $N$ electrons across $m$ subshells with respective degeneracies $g_1, g_2, \dots, g_m$) is crucial for the calculation of atomic structure and spectra, and represents a fundamental issue in statistical physics \cite{Perrot2000,Krief2021,Kurzweil2016,Aberg2024}. For instance, the calculation of hot-plasma radiative opacity, for astrophysical applications or laser-fusion experiments, requires the most complete list of relevant configurations, in order to handle all the radiative one-electron transitions contributing to the photo-absorption cross-section \cite{Pain2023}. However, this is a challenging combinatorial problem (belonging to the class of ``bounded partitions''), and the number of electronic configurations is typically computed numerically through direct multiple summations, which require the evaluation of nested loops. A few years ago, efficient double recursion relations concerning the number of electrons and orbitals were introduced \cite{Gilleron2004}. More recently, we obtained an explicit formula, involving multinomial coefficients and a summation over integer partitions \cite{Pain2020}, as well as a very accurate approximate formula \cite{Pain2025}. However, we could not find an analytical expression in the literature that applies universally. In this paper, we propose an exact analytical formula for the number of atomic configurations, whatever the number of subshells, their degeneracies and the number of electrons. 

The determination of the total angular momentum multiplicities was first investigated by nuclear physicists in the framework of the shell model and later by atomic physicists for electronic configurations. Considering a system of $N$ identical fermions, the problem boils down to deducing the allowed total angular momenta $J$ to which they may couple. Some values of $J$ are forbidden by anti-symmetrization due to the Pauli exclusion principle, some others occur more than once \cite{Condon1959,Sobelman1972,Sobelman1992}. The number $Q(J)$ of levels with angular momentum $J$ is equal to the number of states with projection $M = J$ minus the number of states with $M = J + 1$, i.e., $Q(J)=P(M=J)-P(M=J+1)$ (sometimes referred to as ``the Bethe formula''). The determination of $P(M)$ and $Q(J)$ is therefore important both in nuclear \cite{Zamick2005, Zamick2013, Xin2022} and atomic physics \cite{Katriel1989}. In nuclear physics, they play a major role in the modeling of the $J-$pairing interaction \cite{Zhao2005}. In the present work, we place ourselves within the framework of atomic physics, and therefore consider electrons in atomic configurations. This being said, all the results apply equally well to protons and neutrons, which is why we often use the term ``fermions'', especially when referring to work carried out in a field related to nuclear physics. The problem was addressed by different methods, such as group theory \cite{Cleary1971,Hirst1986}, moment methods \cite{Carlson1989}, trace techniques \cite{Subramanian1974}, Molien functions \cite{Raychev1990}, diagrammatic techniques \cite{Planelles1996} or fractional-parentage coefficients \cite{Pain2019}.

In a previous work \cite{Poirier2021a}, we proposed a statistical analysis of the distribution $P(M)$ in the case of a relativistic configuration (i.e. made of subshells $j^N$). Using the generating-function formalism and the properties of Gaussian binomial coefficients \cite{Comtet1974,Andrews1984,Stanley2011}, we derived recursion relations for $P(M)$ as well as analytical expressions for the cumulants. In the same work, we carried out an analysis of the distribution $P(M)$ using Gram–Charlier and Edgeworth expansion series at any order. However, although $Q(J)$ can be deduced from $P(M)$, only very few properties of $Q(J)$ are explicitly known. Thus, following our analysis of the magnetic quantum number distribution, we proposed an expression for the generating function of the distribution $Q(J)$, from which we derived efficient recurrence relations. Second, using the analysis carried out in Ref. \cite{Poirier2021a}, we generalized the above-mentioned Bethe formula in the form of a Gram–Charlier-like series and discussed its convergence \cite{Poirier2021b}. More recently, we published a general method to obtain the distribution of the total quantum number $M$ for a set of $N$ identical fermions with momentum $j$. The method relied on a recursive procedure on $N$, yielding closed-form expressions, which were found to be linear combinations of piecewise polynomials, according to congruence properties. The calculations were tractable, but the resulting expressions were cumbersome. In the present work, we obtain, using integration properties of the generating functions of the different quantities, the first general explicit formulas for the number of configurations and the distributions $P(M)$ and $Q(J)$. 

The formalism is presented in section \ref{sec2}, and applied to the calculation of the number of configurations. The issue of the distributions of the magnetic moment and total angular momentum of a $j^N$ subshell is addressed in section \ref{sec3}.

\section{Number of configurations: new expression by direct integration}\label{sec2}

The number $\mathscr{N_C}$ of ways to distribute $N$ electrons in an ensemble of $m$ subshells with degeneracies $g_1$, $g_2$, ..., $g_m$ is a cornerstone of configuration generators required by hot-plasma opacity calculations. The brute-force calculation reads
\begin{equation*}
    \mathscr{N_C}=\underbrace{\sum_{p_1=0}^{g_1}\sum_{p_2=0}^{g_2}\cdots\sum_{p_m=0}^{g_m}}_{\sum_{i=1}^mp_i=N}1
\end{equation*}
or equivalently
\begin{equation*}
    \mathscr{N_C}=\sum_{p_1=0}^{g_1}\sum_{p_2=0}^{g_2-p_1}\sum_{p_3=0}^{g_2-p_1-p_2}\cdots\sum_{p_{m-1}=0}^{g_m-p_1-p_2-\cdots-p_{m-2}}H\left(g_m-\left(N-\sum_{i=1}^{m-1}p_i\right)\right),
\end{equation*}
where $H$ is the usual Heaviside function. Actually, $\mathscr{N_C}$ is the coefficient of $x^N$ in the generating function \cite{Wilf2005}:
\begin{equation*}
    g(x)=\prod_{i=1}^m\sum_{j=0}^{g_i}x^j=\prod_{i=1}^m\frac{\left(1-x^{g_i+1}\right)}{\left(1-x\right)}
\end{equation*}
and can be expressed as \cite{Pain2020}:
\begin{equation}\label{A1}
    \mathscr{N_C}=\sum_{i=0}^N\binom{i+m-1}{i}
    \sum_{\vec{\alpha}/\sum_{j=1}^m\alpha_j=N-i}\frac{1}{\alpha_1!\alpha_2!...\alpha_m!}
    \prod_{j=1}^m\left(\delta_{\alpha_j,0}-(g_j+1)!\delta_{\alpha_j,g_j+1}\right).
\end{equation}
If we gather the $n_1$ subshells of degeneracy $g_1$, the $n_2$ subshells of degeneracy $g_2$, ..., the $n_s$ subshells of degeneracy $g_s$ (with therefore $n_1+n_2+\cdots n_s=m$), we obtain \cite{Pain2020}:
\begin{equation}\begin{split}
    \mathscr{N_C}=&\sum_{i_1=0}^{n_1}\sum_{i_2=0}^{n_2}\cdots\sum_{i_s=0}^{n_s}(-1)^{i_1+i_2+\cdots n_s}\binom{n_1}{i_1}\binom{n_2}{i_2}\cdots\binom{n_s}{i_s} \\
              &\times\binom{n_1+\cdots+n_s-1+N-i_1\left(g_1+1\right)-i_2\left(g_2+1\right)-\cdots-i_s\left(g_s+1\right)}{n_1+\cdots+n_s-1}.
\end{split}\end{equation}
We have also
\begin{equation*}
    \mathscr{N_C}=\frac{1}{2\pi i}\oint\frac{1}{z^{N+1}}\prod_{k=1}^m\frac{\left(1-z^{g_k+1}\right)}{\left(1-z\right)}\,\mathrm{d}z,
\end{equation*}
where $\displaystyle \oint$ denotes a contour integral over the unit circle.
A simple expression follows from the parametrization $z=e^{i\theta}$ of the unit circle:
\begin{equation}\label{A2}
    \mathscr{N_C}=\frac{1}{2i\pi}\int_{C}\prod_{k=1}^m\frac{\left(1-z^{g_k+1}\right)}{z^{Ng_k/G}\left(1-z\right)}\,\frac{\mathrm{d}z}z=\frac{1}{2\pi}\int_{-\pi}^{\pi}\prod_{k=1}^m\frac{\left(1-e^{i(g_k+1)\theta}\right)}{e^{iNg_k\theta/G}\left(1-e^{i\theta}\right)}\,\mathrm{d}\theta,
\end{equation}
where $C = \left\{z\in\mathbb{C}, \quad |z|=1\right\}$ and $\displaystyle G=\sum_{k=1}^mg_k$. We have 
\begin{equation*}
    \mathscr{N_C}=\frac{1}{2\pi}\int_{-\pi}^{\pi}f(\theta)\,\mathrm{d}\theta,
\end{equation*}
with
\begin{equation*}
    f(\theta)=\prod_{k=1}^m\frac{\left(1-e^{i(g_k+1)\theta}\right)}{e^{iNg_k\theta/G}\left(1-e^{i\theta}\right)}.
\end{equation*}
Setting $x=(\theta+\pi)/(2\pi)$, one obtains
\begin{equation*}
    \mathscr{N_C}=\int_{0}^{1}\widetilde f(x)\,\mathrm{d}x,
\end{equation*}
with (note that $G$ and $g_k$, $\forall k\in [1,m]$ are always even):
\begin{equation*}
    \widetilde f(x)=\prod_{k=1}^m\frac{\left(1+e^{2i\pi (g_k+1)x}\right)}{e^{iNg_k(2x-1)\pi/G}\left(1+e^{2i\pi x}\right)}.
\end{equation*}
The latter function is a trigonometric polynomial, which can be put in the form
\begin{equation*}
    \widetilde f(x)=\sum_{l=-N}^{d}a_l\,e^{2i\pi lx}
\end{equation*}
and $d=G-N$. Because of the electron-hole symmetry, we can restrict ourselves to $N<G/2$ (and we have $-N>-d$). It is easy to check that
\begin{equation}\label{intf}
    \int_0^1\widetilde f(x)\mathrm{d}x=a_0.
\end{equation}
Let us now calculate
\begin{equation}\label{basis}
    \frac{1}{(d+1)}\sum_{j=0}^{d}\widetilde f\left(\frac{j+1/2}{d+1}\right).
\end{equation}
We have to isolate the case $l=0$:
\begin{equation*}
    \frac{1}{d+1}\sum_{j=0}^{d}\widetilde f\left(\frac{j+1/2}{d+1}\right)=\frac{1}{d+1}\left(\sum_{\substack{l=-N,\\ l\ne 0}}^{d}a_l\,\sum_{j=0}^{d}e^{2i\pi l\frac{(j+1/2)}{d+1}}+(d+1)a_0\right),
\end{equation*}
since, for the $l=0$ case:
\begin{equation*}
    \sum_{j=0}^{d}1=d+1.
\end{equation*}
We have also, for $l\ne 0$:
\begin{equation*}
    \sum_{j=0}^{d}e^{2i\pi l\frac{(j+1/2)}{d+1}}=e^{i\frac{\pi l}{d+1}}\,\frac{1-e^{2i\pi l}}{1-e^{2i\pi l/(d+1)}}=0.
\end{equation*}
We thus have
\begin{equation*}
    \frac{1}{d+1}\sum_{j=0}^{d}\widetilde f\left(\frac{j+1/2}{d+1}\right)=a_0,
\end{equation*}
and using Eq. (\ref{intf}):
\begin{equation*}
    \int_0^1\widetilde f(x)\,\mathrm{d}x=\frac{1}{d+1}\sum_{j=0}^{d}\widetilde f\left(\frac{j+1/2}{d+1}\right).
\end{equation*}
Since $d=G-N$, one gets
\begin{equation*}
    \mathscr{N_C}=\frac{1}{G-N+1}\sum_{j=0}^{G-N}\widetilde f\left(\frac{j+1/2}{G-N+1}\right),
\end{equation*}
i.e.,
\begin{align}\label{resu}
    \mathscr{N_C}=\frac{1}{G-N+1}\sum_{j=0}^{G-N}\prod_{k=1}^m\frac{\left(1+e^{2i\pi (g_k+1)\frac{(j+1/2)}{(G-N+1)}}\right)}{e^{iNg_k\left(2\frac{(j+1/2)}{(G-N+1)}-1\right)\pi/G}\left(1+e^{2i\pi \frac{(j+1/2)}{(G-N+1)}}\right)}.
\end{align}
It is possible to simplify a little bit this expression. Setting $\omega = e^{\frac{2i\pi}{G-N+1}}$, Eq. \eqref{resu} reads
\begin{equation*}
    \mathscr{N_C}=\frac{1}{G-N+1}\sum_{j=0}^{G-N}\prod_{k=1}^m\frac{\left(1+\omega^{(j+1/2)(g_k+1)}\right)}{\omega^{(j+1/2)\frac{Ng_k}G} e^{-i\pi\frac{Ng_k}G}\left(1+\omega^{j+1/2}\right)}.
\end{equation*}
This gives
\begin{equation}\label{finalconf}
    \mathscr{N_C}=\frac{1}{G-N+1}\sum_{j=0}^{G-N}\frac{(-1)^N }{\omega^{N(j+1/2)}\left(1+\omega^{j+1/2}\right)^{m}}\prod_{k=1}^m\left(1+\omega^{(j+1/2)(g_k+1)}\right), \quad \omega = e^{\frac{2i\pi}{G-N+1}}.
\end{equation}
Let us recall that $G$ is the total degeneracy of the ensemble of $m$ subshells (each subshell having a degeneracy $g_k, k\in\{1,m\}$), and $N$ the number of fermions. The set of $\omega^r$ with $r\in\{0,G-N\}$ represents the $(G-N+1)^{th}$ roots of unity. It forms a cyclic group and $\omega^r$ is a generator of this group if and only if $r$ and $G-N+1$ are coprime. These roots are then called primitive $(G-N+1)^{th}$ roots of unity. We do not take advantage of this algebraic property here, but it may lead to further improvements, in connection with cyclotomic polynomials. The choice of $j+1/2$ from the beginning (see Eq. (\ref{basis})), instead of $j$ for instance, enables one to avoid cancellations in the denominator. Indeed, since $j$ is a natural number, the quantity $\omega^{j+1/2}$ can not be equal to $-1$, whatever the value (parity) of the number of electrons $N$.

Let us consider $g_1=2$, $g_2=2$, $g_3=6$, $g_4=2$, $g_5=6$, $g_6=10$, $g_7=2$, $g_8=6$, $g_9=10$, hence, $m=9$ and $G=46$. Expression (\ref{finalconf}) gives 161 for $N=3$ electrons, 459 for $N=4$, 4506 for $N=5$, 13149 for $N=9$ and 116883 for $N=16$.

The new expression (\ref{finalconf}) is very efficient from the numerical point of view, since it requires roughly only
\begin{equation}
    m\times N
\end{equation}
operations, while formulas (\ref{A1}) and (\ref{A2}) are much more expensive, since they involve multiple sums and products of factorials (via binomial or multinomial coefficients), as well as partitions (see the Appendix A of Ref. \cite{Pain2020}). The new formula has exactly the same complexity as the recurrence relation published in Ref. \cite{Gilleron2004}:
\begin{equation*}
    \mathscr{N_C}(N,m)=\sum_{i=0}^{\min(N,g_m)}\mathscr{N_C}(N-i,m-1)
\end{equation*}
with $\mathscr{N_C}(N,m)=\delta(N)$, where we have introduced the dependency of $\mathscr{N_C}$ with respect to the number of electrons $N$ and the number of subshells $m$.

The numbers of configurations for the 10 subshells of the first four $n=1$ to $n=4$ shells (i.e., 1s, 2s, 2p, 3s, 3p, 3d, 4s, 4p, 4d and 4f) and different values of the number of electrons are given in table \ref{tab1}. We can check the electron-hole symmetry is satisfied. The largest number of configurations corresponds to $N=G/2$, where $G$ is the total degeneracy
\begin{equation*}
    2\sum_{n=1}^4n^2=60.
\end{equation*}

\begin{table}
\centering
\begin{tabular}{cc}\hline\hline
Number of electrons & Number of configurations\\\hline\hline
3 & 216\\
5 & 1782\\
10 & 50220\\
15 & 352487\\
20 & 1142430\\
25 & 2167311\\
30 & 2656763\\
35 & 2167311\\
40 & 1142430\\
45 & 352487\\
50 & 5022\\
55 & 1782\\
57 & 216\\\hline\hline
\end{tabular}
\caption{Numbers of configurations for the 10 subshells of the first four $n=1$ to $n=4$ shells (i.e., 1s, 2s, 2p, 3s, 3p, 3d, 4s, 4p, 4d) and different values of the number of electrons.}\label{tab1}
\end{table}

\section{Determination of the number of states with fixed angular momentum projection $M$}\label{sec3}

\subsection{Previous (cumbersome) results: examples for 3, 4 and 5 fermions}\label{subsec31}

Over the past few years, we derived explicit formulas for $P(M)$ and $Q(J)$ involving piecewise polynomials and congruences properties \cite{Poirier2021,Poirier2024} for small numbers of fermions. Although we could also obtain an algorithm to get such expressions whatever the number of fermions, such an approach has two drawbacks: first, it does not provide any explicit formula for any number of fermions, and second, the resulting expressions are rather complicated, even for small numbers of fermions, as illustrated below for 3, 4 and 5 fermions.

\subsubsection{Three-fermion case}\label{subsubsec311} \hfill\\

We found in Ref. \cite{Poirier2021} that, in the three-fermion case ($N=3$) and setting $M=j-q$, with $q$ non-negative integer ($q=1,2\dots J_\text{max}-j$), the distribution of angular-momentum projection can be put in the form:
\begin{equation*}
    P(j-q;j,3)=\frac13\left(j+\frac{q}{2}\right)^2+\alpha(2j+q)
    -H(q)\left[\frac{q^2}{4}+\gamma(q)\right],
\end{equation*}
with
\begin{equation*}
    \gamma(q)=\left(0,-\frac14\right)
\end{equation*}
for $q\bmod2=(0,1)$ if $-2j+3\le q\le j-1/2$, and $\alpha$ defined as
\begin{equation*}
    \alpha(2j-q)=\left(0,-\frac{1}{12},-\frac{1}{3},\frac{1}{4},-\frac{1}{3},-\frac{1}{12}\right) 
\end{equation*}
if $2j-q\bmod6=(0,1,2,3,4,5)$ respectively. One has also
\begin{equation*}
    P(j+q;j,3) = \frac13\left(j-\frac{q}{2}\right)^2+\alpha(2j-q).
\end{equation*}
For instance, one obtains \cite{Poirier2021}:
\begin{equation*}
    P(j;j,3)=\frac{j^2}3\begin{cases} +\frac14 &\text{ if } j=3n+3/2\\
    -\frac{1}{12}=\frac13\left(j^2-\frac14\right)
    &\text{ if } j=3n+1/2 \text{ or }3n+5/2.\end{cases}
\end{equation*}

\subsubsection{Four-fermion case}\label{subsubsec312} \hfill\\

In the four-fermion case ($N=4$), we obtained, setting $M=2j-n$ ($n$ is an integer number) \cite{Poirier2021}:

\begin{align*}
    P(2j-n;j,4)=&\frac{1}{18}\left(j+\frac{n-1}2\right)^3-\left(\frac16-\frac{\pi(n)}{8}\right)\left(j+\frac{n-1}2\right)\\
                &-H(n)\left[f_1\left(\frac{n}2\right)+\xi(n)\right]+\omega(2j+n-1)
\end{align*}
with $H(n)$ the usual Heaviside function, $\pi(n)=n\bmod2$,
\begin{equation*}
    \xi(n)=\left(-\frac{1}{9},-\frac{1}{72},0,-\frac{17}{72},\frac{1}{9},-\frac{1}{8}\right)
\end{equation*}
if $n\bmod6=(0,1,2,3,4,5)$ respectively,
\begin{equation*}
    f_1(n)=\frac29n^3-\frac{n^2}{6}-\frac{n}{6}+\frac19,
\end{equation*}
and
\begin{equation*}
    \omega(2j+n-1)=\left(0,\frac{1}{72},\frac{1}{9},-\frac{1}{8},-\frac{1}{9},\frac{17}{72},0,-\frac{17}{72},\frac{1}{9},\frac{1}{8},-\frac{1}{9},-\frac{1}{72}\right)
\end{equation*}
if $2j+n-1\bmod12=(0,1,2,3,4,5,6,7,8,9,10,11)$ respectively. In this formula $2j-n$ must be non-negative, $n$ may be negative. Explicitly, $n$ must be such that $-2j+6\le n\le2j$.

\subsubsection{Five-fermion case}\label{subsubsec313} \hfill\\

In the five-fermion case ($N=5$), the distribution of $M$ (number of states yielding total angular momentum projection $M$) reads \cite{Poirier2024}:
\begin{align*}
    P(M;j,5)=&\frac{(5j-M)^4}{2880}-\frac{(5j-M)^3}{288}+\frac{(5j-M)^2}{288}+\frac{(5j-M)}{24}
-\pi(5j-M)\frac{(5j-M)}{32}\\
&+\varphi(\mathrm{mod}(5j-M,12))+\frac{\delta(\mathrm{mod}(5j-M,5))}{5}\\
&-H(3j-M)\left[\frac{(3j-M)^4}{576}-\frac{(3j-M)^3}{96}-\frac{(3j-M)^2}{288}+\frac{(3j-M)}{24}\right.\\
&+\left.\pi(3j-M)\frac{(3j-M)}{32}+\eta(\mathrm{mod}(3j-M,12))\right]\\
&+H(j-M)\left[\frac{(j-M)^4}{288}-\frac{(j-M)^3}{144}-\frac{(j-M)^2}{36}+\pi(j-M)\frac{(j-M)}{16}\right.\\
&+\left.\gamma_5(\mathrm{mod}(j-M,6))\right].
\end{align*}
The values for $\varphi,\eta,\gamma_5$ are given by 
\begin{align*}
    \varphi(m)&=\left(-\frac{1}{5},-\frac{31}{2880},-\frac{3}{40},\frac{1}{320},-\frac{4}{45},-\frac{39}{320},-\frac{3}{40},\frac{329}{2880},-\frac15,-\frac{39}{320},\frac{13}{360},\frac{1}{320}\right)\\
              &=-\frac{1}{5}+\frac{\mathrm{mod}(m,2)}{64}+\frac{-\mathrm{mod}(m,3)+2~\mathrm{mod}(-m,3)}{27} +\frac{\mathrm{mod}(m,4)}{16},
\end{align*}
\begin{align*}
    \eta(m)&=\left(0,-\frac{35}{576},-\frac{1}{72},-\frac{3}{64},\frac{1}{9},-\frac{35}{576},-\frac{1}{8},\frac{37}{576},\frac{1}{9},-\frac{11}{64},-\frac{1}{72},\frac{37}{576}\right)\\
           &=\frac{\mathrm{mod}(m,2)}{64}+\frac{\mathrm{mod}(m,3)+\mathrm{mod}(-m,3)}{27}-\frac{\mathrm{mod}(-m,4)}{16},
\end{align*}
and
\begin{align*}
\gamma_5(m)&=\left(0,-\frac{1}{32},\frac{1}{9},-\frac{1}{32},0,\frac{23}{288}\right)\text{ for }m=(0,1,2,3,4,5)\\
           &=-\frac{\mathrm{mod}(m,2)}{32}+\frac{2\mathrm{mod}(m,3)-\mathrm{mod}(-m,3)}{27}.
\end{align*}
Another algorithm was proposed by Sunko and Svrtan \cite{Sunko1985,Sunko1986}. It relies on the use of determinants and the determination of divisors of specific integers. Since only very few details were given in Refs. \cite{Sunko1985,Sunko1986}, we recall the algorithm in \ref{sec4}, and give some proofs of intermediate results, as well as some possible improvement based on prime-number decomposition.

\subsection{New general exact formulas}\label{subsec32}

The distribution $P(M;j,N)$ for a configuration made of a single relativistic subshell $j^N$ ($j$ being half-integer: 1/2, 3/2, 5/2, ...) of degeneracy $2j+1$ is \cite{Poirier2021b}:
\begin{equation}\label{distriM}
    P(M;j,N)=\frac{1}{2\pi i}\oint\frac{z^{-J_{\mathrm{max}}}}{z^{M+1}}\prod_{p=1}^N\frac{\left(1-z^{2j+2-p}\right)}{\left(1-z^p\right)}\,\mathrm{d}z,
\end{equation}
and the distribution $Q(J;j,N)$ can be expressed as
\begin{equation}\label{distriP}
    Q(J;j,N)=\frac{1}{2\pi i}\oint\frac{z^{-J_{\mathrm{max}}}(z-1)}{z^{J+2}}\prod_{p=1}^N\frac{\left(1-z^{2j+2-p}\right)}{\left(1-z^p\right)}\,\mathrm{d}z,
\end{equation}
where 
\begin{equation}\label{Jmax}
    J_{\mathrm{max}}=N(2j+1-N)/2. 
\end{equation}
The technique described in the previous section applies to integrals (\ref{distriM}) and (\ref{distriP}), as shown below. 

\subsubsection{Case of angular momentum projection (or magnetic quantum number) $M$ distribution $P(M;j,N)$}\label{subsubsec321} \hfill\\

\begin{figure}[!ht]
\centering
\includegraphics[scale=0.45]{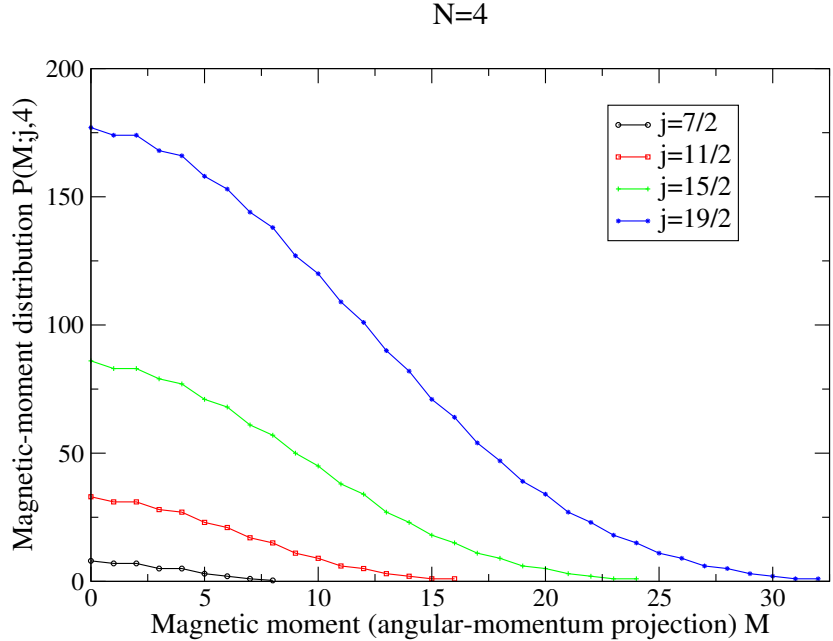}
\caption{Distribution $P(M;j,4)$ for a configuration made of a single $j-$subshell and different values of $j$ (3/2, 7/2, 11/2 and 15/2 respectively), in the case of 4 fermions.}\label{fig1}
\end{figure}

\begin{figure}[!ht]
\centering
\includegraphics[scale=0.45]{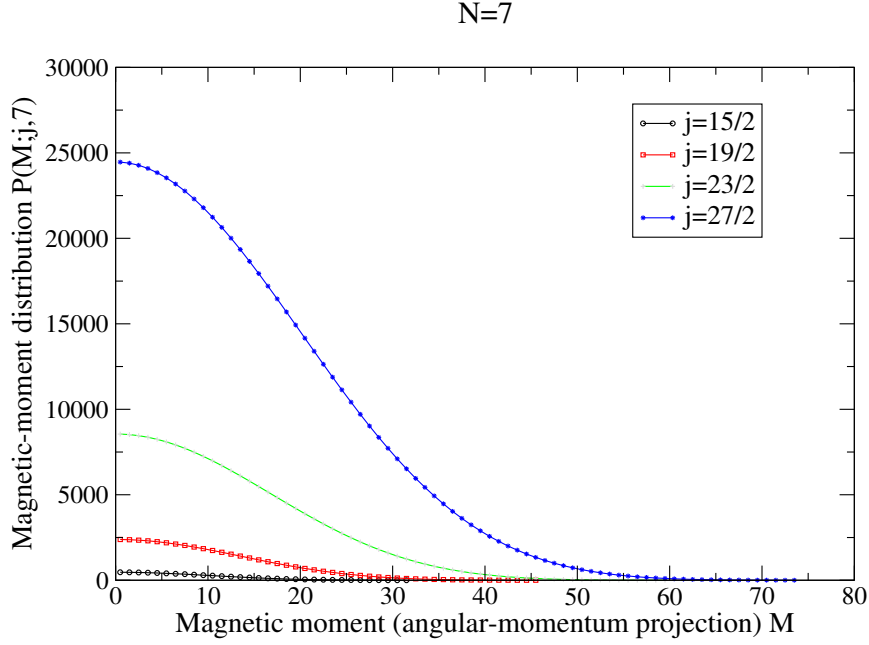}
\caption{Distribution $P(M;j,4)$ for a configuration made of a single $j-$subshell and different values of $j$ (11/2, 15/2, 19/2 and 23/2 respectively), in the case of 7 fermions. The values of the parameters are the same as in Fig. \ref{fig1}.}\label{fig2}
\end{figure}

Making the change of variables $z=e^{i\theta}$ yields
\begin{equation*}
    P(M;j,N)=\frac{1}{2\pi}\int_{-\pi}^{\pi} g(\theta)~\mathrm{d}\theta
\end{equation*}
where
\begin{equation*}
    g(\theta)=e^{-i\left(\frac{N(2j+1-N)}{2}+M\right)\theta}\prod_{p=1}^N\frac{\left(1-e^{i(2j+2-p)\theta}\right)}{\left(1-e^{ip\theta}\right)}
\end{equation*}
and the change of variable $x=(\theta+\pi)/(2\pi)$ gives
\begin{equation*}
    P(M;j,N)=\int_0^1\widetilde{g}(x)~\mathrm{d}x
\end{equation*}
with (since $j=1/2, 3/2, 5/2, 7/2...$ is half-integer):
\begin{equation*}
    \widetilde{g}(x)=e^{-i\left(\frac{N(2j+1-N)}{2}+M\right)\pi(2x-1)}\prod_{p=1}^N\frac{\left(1+(-1)^p~e^{2\pi i(2j+2-p)x}\right)}{\left(1-(-1)^p~e^{2i\pi px}\right)}.
\end{equation*}
The polynomial $\tilde{g}$ can be put in the form
\begin{equation*}
    \widetilde g(x)=\sum_{l=-J_{\mathrm{max}}-M-1}^{J_{\mathrm{max}}-M-1}a_l\,e^{2i\pi lx}
\end{equation*}
and let us set
\begin{equation}\label{dM}
    d_M=\max(J_{\mathrm{max}}-M-1,J_{\mathrm{max}}+M+1).
\end{equation}
Since $P(-M)=P(M)$, we can restrict ourselves to $M\geq 0$ and then $d_M=J_{\mathrm{max}}+M+1$. We get
\begin{equation}\label{resu2}
    P(M;j,N)=\frac{1}{d_M+1}\sum_{r=0}^{d_M}\widetilde{g}\left(\frac{r}{d_M+1}\right).
\end{equation}
It is possible to simplify a little bit the latter expression. Setting $\omega = e^{\frac{2i\pi}{d_M+1}}$, Eq. \eqref{resu2} reads
\begin{equation}\label{PMfin}
    P(M;j,N)=\frac{(-1)^{J_{\mathrm{max}}+M}}{d_M+1}\sum_{r=0}^{d_M}\omega^{-r(J_{\mathrm{max}}+M)}\prod_{k=1}^N\frac{\left(1+(-1)^k~\omega^{r(2j+2-k)}\right)}{\left(1-(-1)^k~\omega^{kr}\right)},
\end{equation}
where $J_{\mathrm{max}}$ is given by Eq. (\ref{Jmax}) and $d_M$ by Eq. (\ref{dM}). In order to avoid cases where the denominator vanishes, one just has, as for the number of configurations, to add $\zeta$ to $r$ in Eqs. (\ref{resu2}) and (\ref{PMfin}), so that $k\zeta$ is never an integer (one can choose for instance $\zeta=1/(N+1)$). We take therefore
\begin{equation*}
    P(M;j,N)=\frac{(-1)^{J_{\mathrm{max}}+M}}{d_M+1}\sum_{r=0}^{d_M}\omega^{-\left(r+\frac{1}{N+1}\right)(J_{\mathrm{max}}+M)}\prod_{k=1}^N\frac{\left(1+(-1)^k~\omega^{\left(r+\frac{1}{N+1}\right)(2j+2-k)}\right)}{\left(1-(-1)^k~\omega^{k\left(r+\frac{1}{N+1}\right)}\right)}.
\end{equation*}
The distribution $P(M;j,4)$ is represented for different values of $J$ (7/2, 11/2, 15/2 and 19/2) in Fig. \ref{fig1}. The first value of $M$ is 0, since the number of electrons is even. The distribution $P(M;j,7)$ is displayed in Fig. \ref{fig2} for even higher values of $j$ (11/2, 15/2, 19/2 and 23/2). Since the distribution $P(M;j,N)$ is symmetrical with respect to the $y$ axis, only positive values of $M$ are represented. In that case, the number of fermions being odd, the minimum value of $M$ is 1/2. We can see that the values of $P(M;j,N)$ vary strongly with $j$.

\subsubsection{Case of angular momentum $J$ distribution $Q(J;j,N)$}\label{subsubsec322} \hfill\\

Making the change of variables $z=e^{i\theta}$ yields
\begin{equation*}
    Q(J;j,N)=\frac{1}{2\pi}\int_{-\pi}^{\pi} h(\theta)~\mathrm{d}\theta
\end{equation*}
where
\begin{equation*}
    h(\theta)=e^{-i\left(\frac{N(2j+1-N)}{2}+J+1\right)\theta}(e^{i\theta}-1)\prod_{p=1}^N\frac{\left(1-e^{i(2j+2-p)\theta}\right)}{\left(1-e^{ip\theta}\right)}
\end{equation*}
and the change of variable $x=(\theta+\pi)/(2\pi)$ gives
\begin{equation*}
    Q(J;j,N)=\int_0^1\widetilde{h}(x)~\mathrm{d}x
\end{equation*}
with
\begin{equation*}
    \widetilde{h}(x)=-e^{-i\left(\frac{N(2j+1-N)}{2}+J+1\right)\pi(2x-1)}(e^{2\pi ix}+1)\prod_{p=1}^N\frac{\left(1+(-1)^p~e^{2\pi i(2j+2-p)x}\right)}{\left(1-(-1)^p~e^{2i\pi px}\right)}.
\end{equation*}
The polynomial $\widetilde{h}(x)$ can be put in the form
\begin{equation*}
    \widetilde h(x)=\sum_{l=-J_{\mathrm{max}}-J-1}^{J_{\mathrm{max}}-J-1}a_l\,e^{2i\pi lx}
\end{equation*}
and let us set
\begin{equation*}
    d_J=\max(-J_{\mathrm{max}}-J-1,J_{\mathrm{max}}+J+1).
\end{equation*}
Therefore, we get
\begin{equation*}
    Q(J;j,N)=\frac{1}{d_J+1}\sum_{r=0}^{d_J}\widetilde{h}\left(\frac{r}{d_J+1}\right).
\end{equation*}
It is possible to simplify a little bit the latter expression. Setting $\omega = e^{\frac{2i\pi}{d_J+1}}$, Eq. \eqref{resu2} reads
\begin{equation}\label{QMfin}
    Q(J;j,N)=-\frac{(-1)^{J_{\mathrm{max}}+J}}{d_J+1}\sum_{r=0}^{d_J}\frac{(1+\omega)}{\omega^{r(J_{\mathrm{max}}+J)}}\prod_{k=1}^N\frac{\left(1+(-1)^k~\omega^{r(2j+2-k)}\right)}{\left(1-(-1)^k~\omega^{kr}\right)}.
\end{equation}
Here also, we add $\zeta=1/(N+1)$ to $r$ and obtain
\begin{equation*}
    Q(J;j,N)=-\frac{(-1)^{J_{\mathrm{max}}+J}}{d_J+1}\sum_{r=0}^{d_J}\frac{(1+\omega)}{\omega^{\left(r+\frac{1}{N+1}\right)(J_{\mathrm{max}}+J)}}\prod_{k=1}^N\frac{\left(1+(-1)^k~\omega^{\left(r+\frac{1}{N+1}\right)(2j+2-k)}\right)}{\left(1-(-1)^k~\omega^{k\left(r+\frac{1}{N+1}\right)}\right)}.
\end{equation*}

\begin{figure}[!ht]
\centering
\includegraphics[scale=0.45]{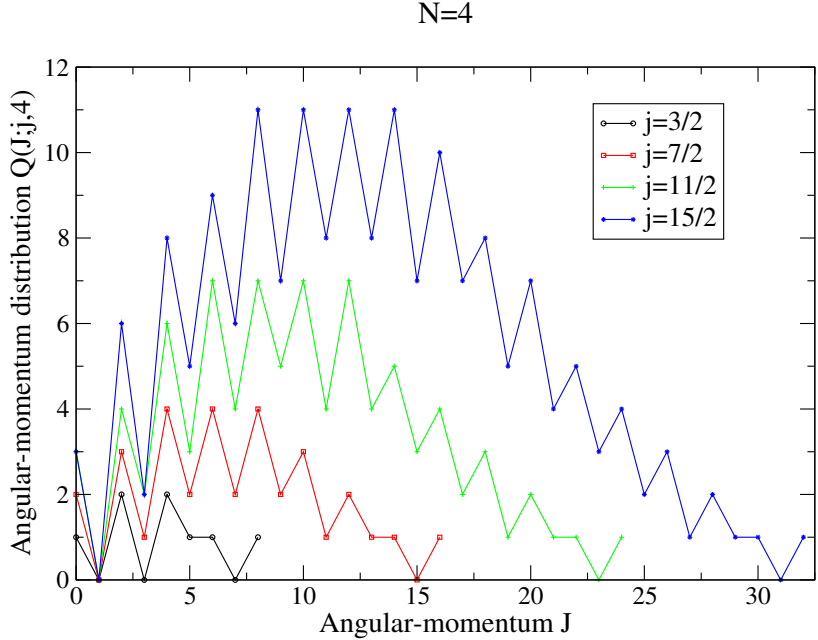}
\caption{Distribution $Q(J;j,4)$ for a configuration made of a single $j-$subshell and different values of $j$, in the case of 4 fermions. The values of the parameters are the same as in Fig. \ref{fig1}.}\label{fig3}
\end{figure}

\begin{figure}[!ht]
\centering
\includegraphics[scale=0.45]{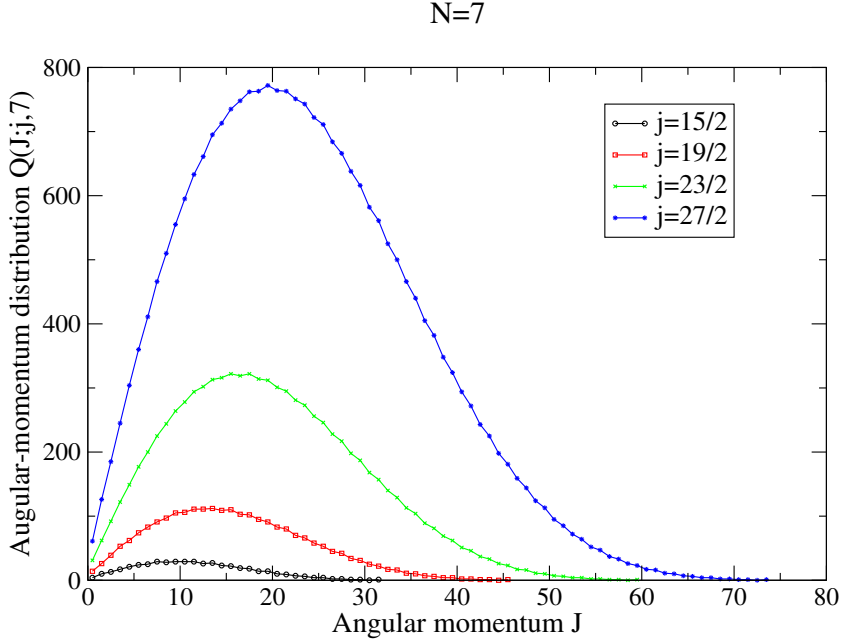}
\caption{Distribution $Q(J;j,4)$ for a configuration made of a single $j-$subshell and different values of $j$, in the case of 7 fermions. The values of the parameters are the same as in Fig. \ref{fig1}.}\label{fig4}
\end{figure}

The distribution $Q(J;j,4)$ is represented for different values of $J$ (7/2, 11/2, 15/2 and 19/2) in Fig. \ref{fig3}. This is the same configuration as in Fig. \ref{fig1}. Similarly, the distribution $Q(J;j,7)$, corresponding to the case of Fig. \ref{fig2}, is displayed in Fig. \ref{fig4}. The distribution $Q(J;j,N)$, on the contrary to $P(M;j,N)$, is asymmetrical.

\subsection{Numerical cost}\label{subsec33}

Let us compare the numerical cost of the new formula with the one of other methods. To this respect we need to estimate the number of operations required to obtain the whole set of $P(M)$ values in a $j^N$ relativistic subshell. The brute force technique consists in evaluating all the 
\begin{equation}\label{eq:nop_bf}
    \binom{2j+1}{N}
\end{equation}
$N$-tuple elements and compute the sum $\sum_{i=1}^N m_i$ for each of them ($-j\leq m_i\leq j$ is the angular-momentum projection of the one-electron state $i$). A more efficient alternative is provided by the recurrence method of Ref. \cite{Gilleron2009}:
\begin{equation}\label{GP}
    P(M;k,n)=P(M;k-1,n)+P(M-m_k;k-1,n-1)
\end{equation}
with $P_{0;k,n}=\delta(M)$, $\forall\, k$, $k=1, ..., N$, $n=1, ..., N$ and $M$ varying from $-J_{\mathrm{max}}$ to $J_{\mathrm{max}}$. This amounts to perform
\begin{equation}\label{eq:nop_recGP}
    N(2j+1)\big(N(2j+1-N)+1\big)
\end{equation}
operations. As a third option, the recurrence over $N$ gives \cite{Poirier2021a}:

\begin{align*}
    &P(n-N-N(2j+1-N)/2;j,N) - P(n-N(2j+1-N)/2;j,N)\nonumber\\
          &\;\;\;\;\;\;\;\;=P(n-2j-2+N-(N-1)(2j+2-N)/2;j,N-1)\nonumber\\
          &\;\;\;\;\;\;\;\;\;\;\;\;\;\;\;\;-P(n-(N-1)(2j+2-N)/2;j,N-1).
\end{align*}
which, with the definition 
\begin{equation*}
    \mathscr{P}_{j,N}(n)=P(n-N(2j+1-N)/2;j,N)
\end{equation*}
with $n$ an integer in the range $0\le n\le N(2j+1-N)$, can be recast into
\begin{equation}\label{eq:recOPM_overN}
    \mathscr{P}_{j,N}(n)=\mathscr{P}_{j,N-1}(n)-\mathscr{P}_{j,N-1}(n-2j-2+N)+\mathscr{P}_{j,N}(n-N).
\end{equation}
This induction relation is initialized by the $N=0$ value and then applied for every $-J_\text{max}(j,\nu)\le M\le J_\text{max}(j,\nu)$ for $1\le \nu\le N$ with $J_\text{max}(j,\nu)=\nu(2j+1-\nu)/2$. Since the formula expresses $P(M;...)$ as a function of three other values of $P$, the number of needed operations reads 
\begin{equation}\label{eq:nop_recN}
3\sum_{\nu=1}^N \left(\nu(2j+1-\nu)+1\right)
 = N\left(3j(N+1)-N^2+4\right).
\end{equation}
This is actually an overestimate since in some cases the recurrence formula involves less than three terms in its right-hand side, due to the selection rules. Moreover, the symmetry property $P(-M;...)=P(M;...)$ is not used. The induction relation \cite{Poirier2021a}:
\begin{equation}\label{eq:recOPM_overj}
    \mathscr{P}_{j,N}(n) = \mathscr{P}_{j-1/2,N}(n)-\mathscr{P}_{j-1/2,N}(n-2j-1) +\mathscr{P}_{j,N}(n-2j-1+N).
\end{equation}
will be initialized with the minimum value $j=(N-1)/2$. If $i$ is twice the iterated 
angular momentum, ranging from $N$ to $2j$, the number of required operations is
\begin{equation}\label{eq:nop_recj}
    3\sum_{i=N}^{2j} \left(N(i+1-N)+1\right) = \frac{3}{2}(2j+1-N)\left(N(2j+1-N)+N+2\right).
\end{equation}
Finally, for our formula (\ref{PMfin}), we have
\begin{align}\label{costPMfin}
    N\times\sum_{M=0}^{J_{\mathrm{max}}}d_M=&N\times \sum_{M=0}^{J_{\mathrm{max}}}(J_{\mathrm{max}}+M+1)\nonumber\\
    =&N\left(\frac{N(2j+1-N)}{2}+1\right)\left(\frac{3N(2j+1-N)}{4}+1\right)
\end{align}
operations. Some examples for the numbers of operations (\ref{eq:nop_bf}), (\ref{eq:nop_recGP}), (\ref{eq:nop_recN}) and (\ref{eq:nop_recj}) are given in Table \ref{tab2}, in the case of an half-filled subshell which leads to the maximum complexity, and compared to the new formula (Eq. (\ref{PMfin})), denoted by ``This work'' in table \ref{tab2}. It may be noted that the recurrence on $j$ (\ref{eq:recOPM_overj}), though using ``unphysical'' quantities, is sometimes more efficient than the recurrence on $N$. The new formula is less efficient than the recurrence relations, but is still much less expensive than the brute-force calculation (denoted by ``Combinatorics'' in table \ref{tab2}).

It is worth making a remark on numerical efficiency. As explained above, we have used the number of integration points necessary for the formula to be exact. But if we use fewer points, we get an approximation of the number of configurations. Thus, if we control the integration error (which requires controlling the derivative of the function we are integrating), we can ensure that the result is an approximation of the exact value. In particular, we can reduce the number of points so that, to the decimal part, we have the correct value. This may reduce the numerical cost. Another alternative would be to resort to a higher-order integration method, which would require far fewer points and therefore a lower computational cost. This would also yield an approximate value, which we hope would be close to an integer — the expected result.

\begin{table}[htbp]
\centering
\begin{tabular}{ccccccccc}
\hline\hline
$j$           & 1/2 & 3/2 & 7/2 & 11/2 & 15/2 & 19/2 & 23/2 & 27/2\\
\hline
Combinatorics & 2 & 6 & 70 & 924 & 12870 & 184756 & 2704156 & 40116600 \\
Recurrence (\ref{GP}) & 4 & 24 & 544 & 2664 & 8320 & 20200 & 41760 & 77224 \\
Recurrence (\ref{eq:recOPM_overN}) & 6 & 27 & 162 & 501 & 1140 & 2175 & 3702 & 5817 \\
Recurrence (\ref{eq:recOPM_overj}) & 6 & 24 & 132 & 396 & 888 & 1680 & 2844 & 4452 \\
This work (\ref{PMfin}) & 3 & 24 & 468 & 3192 & 12936 & 38760 & 95484 & 205128 \\\hline\hline
\end{tabular}
\caption{Number of operations needed to obtain the $P(M)$ distribution 
for the $j^N$ configuration with $N=j+1/2$, using a brute-force technique 
or recurrence relations. Numbers are given according to formulas 
(\ref{eq:nop_bf}), (\ref{eq:nop_recGP}), (\ref{eq:nop_recN}), (\ref{eq:nop_recj}) and (\ref{costPMfin}).}\label{tab2}
\end{table}

\subsection{Total number of $J$ levels}\label{subsec34}

It is worth mentioning that the total number of $J$ levels is given by
\begin{equation*}
\sum_{J=J_\text{min}}^{J_\text{max}}Q(J;j,N)=P\left(J_{\text{min}};j,N\right)
\end{equation*}
where $J_\text{min}=0$ (resp. 1/2) for $N$ even (resp. odd). It was shown in Ref. \cite{Pain2019} that (we do not write the additional $\zeta=1/4$ term inserted in order to avoid the division by zero):
\begin{equation}\label{qtotj3}
    Q_{\mathrm{tot}}\left(j^3\right)=P\left(\frac12;j,3\right)
 =\sum_{i=1/2}^j(i-1/2)=\sum_{t=0}^{j-1/2}t=\frac12\left(j^2-\frac14\right).
\end{equation}
Formula (\ref{PMfin}) gives, for $J=1/2$ and $N=3$, setting $\omega=e^{2i\pi/(2j+3/2)}$:
\begin{equation*}
    Q_{\mathrm{tot}}\left(j^3\right)=P(1/2;j,3)=\frac{(-1)^{3j-\frac{5}{2}}}{3j-\frac{1}{2}}\sum_{r=0}^{3(j-\frac{1}{2})}\omega^{-r(3j-\frac{5}{2})}\prod_{k=1}^3\frac{\left(1+(-1)^k\,\omega^{r(2j+2-k)}\right)}{\left(1-(-1)^k\,\omega^{kr}\right)},
\end{equation*}
since
\begin{equation*}
    J_{\mathrm{max}}=\frac{3}{2}(2j+1-3)=3(j-1)
\end{equation*}
and $d_{1/2}=3(j-1/2)$. For $j=9/2$, one obtains that the total number of levels is equal to 10, which is the right result, as can be checked from Eq. (\ref{qtotj3}). Of course, in that very special (simple) case the new expression is more complicated than (\ref{qtotj3}), but for other values of $M$ (in the general case), the new expression is simpler, as we have seen in section \ref{subsec31}.
We found in Ref. \cite{Poirier2021} that the total number of levels for four fermions of spin $j$ reads (here again we do not write the additional $\zeta=1/5$ term inserted in order to avoid the division by zero):
\begin{equation}\label{qtotj4}
Q_{\mathrm{tot}}\left(j^4\right)=P(0;j,4)= \frac{2}{9}j^3-\frac{j^2}{6}+\frac{j}{6}
 \begin{cases}
 -5/72&\text{ if }j-1/2\bmod3=0,\\
 +3/8&\text{ if }j-1/2\bmod3=1,\\
 +11/72&\text{ if }j-1/2\bmod3=2.
 \end{cases}
\end{equation}
Formula (\ref{PMfin}) gives, for $J=0$ and $N=4$:
\begin{equation*}
    Q_{\mathrm{tot}}\left(j^4\right)=P(0;j,4)=\frac{1}{4(j+1)}\sum_{r=0}^{4j-5}\omega^{-2r(2j-3)}\prod_{k=1}^4\frac{\left(1+(-1)^k\omega^{r(2j+2-k)}\right)}{\left(1-(-1)^k\omega^{kr}\right)},
\end{equation*}
since
\begin{equation*}
    J_{\mathrm{max}}=\frac{4}{2}(2j+1-4)=2(2j-3)
\end{equation*}
and $d_0=4j-5$. As a simple test case, for $j=9/2$, one obtains that the total number of levels is equal to 12, which is the right result, as can be checked from Eq. (\ref{qtotj4}). Here also, in that case the new expression is more cumbersome than (\ref{qtotj4}), but for other values of $M$, the new expression is simpler, as we have seen in section \ref{subsec31}.

\section{Conclusion}\label{sec5}

In this work, we have presented, for the first time, exact explicit formulas for the number of atomic configurations and for the distributions of the total magnetic quantum number $M$ and of the total angular momentum $J$. The corresponding relations are based on the idea that since the generating functions of the latter quantities are trigonometric polynomials, their integral can be evaluated exactly as a finite sum on specific points, given by roots of unity, with a small shift in their power in order to ease the numerical implementation in a computer algebra system for instance. The new expressions are much more compact than the ones previously published. Indeed, the latter were limited to small number of fermions (less than or equal to 6) and were rather complicated and lengthy, involving piece-wise polynomials and congruences. Moreover, only a general procedure was provided for any number of fermions. The present formulas can be applied in any case, are very simple and easy to implement. Although the expressions presented here are less efficient numerically than recursion relations (except for the number of configurations, for which the complexities are equivalent), they should benefit to all the computations of atomic structure and spectra, as well as to the statistical modeling of arrays of absorption or emission lines in hot plasmas. For instance, an exact analytical formula can be obtained for the number of lines $\mathrm{L}_{C-C'}$ an array of electric-dipole lines between two configurations $C$ and $C'$, by inserting expression (\ref{QMfin}) into
\begin{equation*}
    \mathrm{L}_{C-C'}=\sum_{J=J_{\mathrm{min}}}^{J_{\mathrm{max}}}Q_C(J)\left[Q_{C'}(J+1)+Q_{C'}(J-1)+Q_{C'}(J)\right],
\end{equation*}
where $J=0$ is discarded in the last term. Moreover, a very accurate estimate of the number of configurations can be obtained, while reducing the number of points in the integration. In the future, we plan to derive similar expressions as the ones presented here, but in the framework of spin-adapted spaces \cite{Lain1998}. The techniques presented here can also be of interest for quantum-chemistry calculations, using software packages {\it Gaussian} \cite{Gaussian} or {\it ORCA} \cite{ORCA} for instance, which feature a wide variety of methods ranging from semi-empirical methods to density functional theory to correlated single- and multi-reference wave function-based methods. When dealing with molecules, however, one has to take into account other quantum numbers, and discuss the relative importance of rotation, vibration, translation and electronic transitions, which determines the coupling scheme.

In addition, generalization of the present work to bosonic systems would be worth considering, for instance in the context of the Landau–Yang theorem \cite{Yang1950,Landau1948}, which states that a massive particle with spin 1 cannot decay into two photons. Such a theorem has important consequences in particle physics (the Z boson cannot decay into two photons, the Higgs boson cannot have spin 1, \emph{etc.}). Furthermore, the techniques presented here can be applied for bosons as well in nuclear-shell theory \cite{Zhao2003,Bao2016}. 

\appendix

\section{Alternative approach}\label{sec4}

It is well known that $P(M)$ is actually the coefficient of $q^M$ in
\begin{equation*}
    q^{-J_{\mathrm{max}}}\,\qbin{2j+1}{N}=q^{-\frac{N(2j+1-N)}{2}}\,\qbin{2j+1}{N}
\end{equation*}
where
\begin{equation*}
    \qbin{p}{r}=\frac{(1-q^p)(1-q^{p-1})\cdots(1-q^{p-r+1})}{(1-q)(1-q^2)\cdots(1-q^r)}
\end{equation*}
represents the $q$-binomial coefficient (also referred to as ``Gaussian polynomial'').
Sunko and Svrtan suggested to write (removing the $q^{-J_{\mathrm{max}}}$ term and replacing $P(M)$ by $\mathscr{P}(q)$) \cite{Sunko1985,Sunko1986}:
\begin{equation*}
    \mathscr{R}(q)=\sum_{m\ge 0}c_mq^m.
\end{equation*}
Let us write
\begin{equation*}
    \mathscr{R}(q)=\exp\left[\psi(q)\right],
\end{equation*}
with
\begin{equation*}
    \psi(q)=\sum_{k\geq 1}\frac{1}{k}\eta_kq^k.
\end{equation*}
It turns out that $c_m$ and $\eta_k$ are connected by a determinantal identity \cite{Faa}. It is clear that
\begin{equation*}
\mathscr{R}'(q)=\psi'(q)\,\mathscr{R}(q),
\end{equation*}
which reads
\begin{equation*}
    \sum_{m\geq 1}m\,c_m\,q^{m-1}=\left(\sum_{k\geq 1}\eta_kq^{k-1}\right)\left(\sum_{m\ge 0}c_mq^m\right)
\end{equation*}
or equivalently
\begin{equation*}
    \sum_{m\geq 0}(m+1)\,c_{m+1}\,q^{m}=\left(\sum_{k\geq 0}\eta_{k+1}q^{k}\right)\left(\sum_{j\ge 0}c_jq^j\right)
\end{equation*}
yielding
\begin{equation*}
    mc_m=\eta_1c_{m}+\eta_2c_{m-1}+\cdots+\eta_{m-1}c_1+\eta_m,
\end{equation*}
or also
\begin{equation}\label{26}
    (m+1)c_{m+1}=\sum_{k=1}^m\eta_{k+1}c_{m-k}.
\end{equation}
The latter recurrence relation can be put in the form
\begin{equation}\label{25}
    c_m=\frac{1}{m!}\left|
\begin{array}{ccccccc}
\eta_1     & -1         & 0          & 0      & 0      & \cdots & 0 \\
\eta_2     & \eta_1     & -2         & 0      & 0      & \cdots & 0 \\
\eta_3     & \eta_2     & \eta_1     & -3     & 0      & \cdots & 0 \\
\vdots     & \vdots     & \vdots     & \ddots & \ddots & \ddots & \vdots \\
\vdots     & \vdots     & \vdots     & \vdots & \eta_1 & (-m+2) & 0 \\
\eta_{m-1} & \eta_{m-2} & \eta_{m-3} & \cdots & \eta_2 & \eta_1 & (-m+1) \\
\eta_m     & \eta_{m-1} & \eta_{m-2} & \cdots & \eta_3 & \eta_2 & \eta_1 \\
\end{array}
\right|.
\end{equation}
This can be proven by induction. For $m=1$, since $c_0=1$, Eq. (\ref{25}) gives $c_1=\eta_1$, which is the case in Eq. (\ref{26}). Let us now expand the determinant (\ref{25}) with respect to the last line; we get
\begin{equation*}
    \frac{1}{m!}\left|
\begin{array}{ccccccc}
\eta_1     & -1         & 0          & 0      & 0      & \cdots & 0 \\
\eta_2     & \eta_1     & -2         & 0      & 0      & \cdots & 0 \\
\eta_3     & \eta_2     & \eta_1     & -3     & 0      & \cdots & 0 \\
\vdots     & \vdots     & \vdots     & \ddots & \ddots & \ddots & \vdots \\
\vdots     & \vdots     & \vdots     & \vdots & \eta_1 & (-m+2) & 0 \\
\eta_{m-1} & \eta_{m-2} & \eta_{m-3} & \cdots & \eta_2 & \eta_1 & (-m+1) \\
\eta_m     & \eta_{m-1} & \eta_{m-2} & \cdots & \eta_3 & \eta_2 & \eta_1 \\
\end{array}
\right|=\frac{1}{m!}\sum_{j=0}^{m-1}\eta_{m-j}\Delta_j,
\end{equation*}
where
\begin{equation*}
    \Delta_j=(-1)^{m+j}\left|
\begin{array}{cccccccccc}
\eta_1     & -1         & 0      & 0      & 0      & \cdots & \cdots & \cdots & \cdots       & 0 \\
\eta_2     & \eta_1     & -2     & 0      & 0      & \cdots & \cdots & \cdots & \cdots       & \vdots \\
\eta_3     & \eta_2     & \eta_1 & -3     & 0      & \cdots & \cdots & \cdots & \cdots       & \vdots \\
\vdots     & \vdots     & \vdots & \ddots & \ddots & \ddots & \cdots & \cdots & \cdots       & \vdots \\
\vdots     & \vdots     & \vdots & \vdots & \eta_1 & (-j+1) & \ddots & \ddots & \cdots       & \vdots \\
\vdots     & \vdots     & \vdots & \vdots & \vdots & \eta_1 & (-j-1) & \ddots & \cdots       & \vdots \\
\vdots     & \vdots     & \vdots & \vdots & \vdots & \vdots & \eta_1 & \ddots & \ddots       & \vdots \\
\vdots     & \vdots     & \vdots & \vdots & \vdots & \vdots & \vdots & \eta_1 & \ddots       & \vdots \\
\vdots     & \vdots     & \vdots & \vdots & \vdots & \vdots & \vdots & \vdots & \ddots       & 0 \\
\eta_{m-1} & \eta_{m-2} & \cdots & \cdots & \cdots & \eta_{j-2} & \cdots & \cdots & \qquad\eta_1 & \;\;\;(-m+1) \\
\end{array}
\right|
\end{equation*}
giving
\begin{align*}
    \Delta_j=&(-1)^{m+j}\left[(-j-1)(-j-2)(-j-3)\times\cdots\times(-m+1)\right]\\
    &\times\left|\begin{array}{cccccc}
\eta_1     & -1         & 0          & 0      & \cdots & 0 \\
\eta_2     & \eta_1     & -2         & 0      &        & \cdots \\
\eta_3     & \eta_2     & \eta_1     & -3     & 0      & \cdots \\
\vdots     & \vdots     & \vdots     & \vdots & \vdots & \vdots \\
\vdots     & \vdots     & \vdots     & \vdots & \eta_1 & (-j+1) \\
\eta_{j}   & \eta_{j-1} & \eta_{j-2} & \cdots & \eta_2 & \eta_1 \\
\end{array}
\right|.
\end{align*}
The induction assumption is: $\forall j\leq m-1$,
\begin{equation*}
\left|\begin{array}{ccccccc}
\eta_1     & -1         & 0          & 0      & 0       & \cdots \\
\eta_2     & \eta_1     & -2         & 0      &         & \cdots \\
\eta_3     & \eta_2     & \eta_1     & -3     & 0       & \cdots \\
\vdots     & \vdots     & \vdots     & \vdots & \vdots  & \vdots \\
\vdots     & \vdots     & \vdots     & \vdots & \eta_1  & (-j+1) \\
\eta_{j}   & \eta_{j-1} & \eta_{j-3} & \cdots & \eta_2  & \eta_1 \\
\end{array}
\right|=j!\,c_j,
\end{equation*}
and therefore, since
\begin{equation*}
    -(j+1)(-j-2)(-j-3)\times\cdots\times(-m+1)=(-1)^{m-j}\frac{(m-1)!}{j!},
\end{equation*}
one gets
\begin{align*}
\left|\begin{array}{ccccccc}
\eta_1     & -1         & 0          & 0      & 0       & \cdots \\
\eta_2     & \eta_1     & -2         & 0      &         & \cdots \\
\eta_3     & \eta_2     & \eta_1     & -3     & 0       & \cdots \\
\vdots     & \vdots     & \vdots     & \vdots & \vdots  & \vdots \\
\vdots     & \vdots     & \vdots     & \vdots & \eta_1  & (-m+1) \\
\eta_{m}   & \eta_{m-1} & \eta_{m-3} & \cdots & \eta_2  & \eta_1 \\
\end{array}
\right|=&\sum_{j=0}^{m-1}\frac{1}{m!}\,\eta_{m-j}(-1)^{2m+j-j}(m-1)!c_j\\
=&\frac{1}{m}\sum_{j=0}^{m-1}\eta_{m-j}\,c_j,
\end{align*}
which is exactly Eq. (\ref{26}) and completes the proof.

The point is that it is easy to obtain a closed-form expression for the $\eta_k$'s of a Gaussian polynomial because its logarithm is a simple expression:
\begin{align*}
    \ln\qbin{p}{r}=&\ln(1-q^p)+\ln(1-q^{p-1})+\cdots+\ln(1-q^{p-r+1})\\
    =&-\ln(1-q)-\ln(1-q^2)-\cdots-\ln(1-q^r).
\end{align*}
Using the series expansion of $\ln(1-x)$, one gets
\begin{align*}
\ln\qbin{p}{r}=&-q^p-\frac{q^{2p}}{2}-\cdots-\frac{q^{np}}{n}-\cdots\\
    &-q^{p-r+1}-\frac{q^{2(p-r+1)}}{2}-\cdots-\frac{q^{n(p-r+1)}}{n}-\cdots\\
    &+q+\frac{q^2}{2}\cdots+\cdots+\frac{q^n}{n}+\cdots\\
    &+q^r+\frac{q^{2r}}{2}\cdots+\cdots+\frac{q^{nr}}{n}+\cdots
\end{align*}
and thus
\begin{equation}\label{PMdiv}
    \eta_m=\sum_{s=1, s|m}^{\min(r,p-r)}s-\sum_{s=\max(r,p-r)+1, s|m}^{p}s,
\end{equation}
where $s|m$ means that $s$ divides $m$, and we have 
\begin{equation*}
    c_m=\frac{1}{m}\left(\eta_m+\sum_{q=1}^{m-1}c_q\eta_{m-q}\right)
\end{equation*}
with $c_0=1$ and $c_k=0$ for $k<0$. Let us introduce the decomposition of $m$ in prime factors
\begin{equation*}
    m=\prod_{i=1}^tp_{i}^{a_i}
\end{equation*}
where $t$ is the number of distinct prime factors of $m$, $p_i$ is the $i^{th}$ prime factor, and $a_i$ is the maximum power of $p_i$ by which $m$ is divisible, then we have: 
\begin{equation*}
    \sigma(m)=\prod_{i=1}^{t}\sum_{j=0}^{a_{i}}p_{i}^{j}=\prod_{i=1}^{t}\left(1+p_{i}+p_{i}^{2}+\cdots +p_{i}^{a_{i}}\right).
\end{equation*}
which is equivalent to 
\begin{equation*}
    \sigma(m)=\prod _{i=1}^{t}{\frac {p_{i}^{(a_{i}+1)}-1}{p_{i}-1}}.
\end{equation*}
Here, the problem is slightly more complicated; indeed, we need to calculate (see Eq. (\ref{PMdiv})) the sum of divisors of $m$ lying between two natural numbers $n_0$ and $n_1$ (both between 1 and $m$ and $n_1\geq n_0$). This means that we have to consider only the divisors of the kind
\begin{equation*}
    s=p_1^{\beta_1}\,p_2^{\beta_2}\cdots p_t^{\beta_t}
\end{equation*}
with $\forall j\in [1,t]$, $0\leq \beta_j\leq a_j$, but with the constraint(s):
\begin{equation*}
    n_0\leq p_1^{\beta_1}\,p_2^{\beta_2}\cdots p_t^{\beta_t}\leq n_1.
\end{equation*}
Knowing the prime-factor decomposition of $n_0$ and $n_1$: 
\begin{equation*}
    n_0=\prod_{i=1}^tp_{i}^{\gamma_i}
\end{equation*}
and
\begin{equation*}
    n_1=\prod_{i=1}^tp_{i}^{\delta_i},
\end{equation*}
we get
\begin{equation*}
    \eta_m=\prod _{i=1}^{t}p_i^{\gamma_i}\frac {p_{i}^{(\delta_{i}-\gamma_i+1)}-1}{p_{i}-1}=\prod _{i=1}^{t}\frac {p_{i}^{(\delta_{i}+1)}-p_i^{\gamma_i}}{p_{i}-1}.
\end{equation*}
This approach is worth mentioning, since it provides a general algorithm to determine $\mathscr{R}(q)$, and thus $P(M)$, but does not lend itself to an explicit formula.


\begin{thebibliography}{99}

\bibitem{Perrot2000} Perrot F and Blenski T 2000 {\it J. Phys. IV France} {\bf 10}, Pr5–473--Pr5–480

\bibitem{Krief2021} Krief M 2021 {\it Phys. Rev. E} {\bf 103} 033206

\bibitem{Kurzweil2016} Kurzweil Y and Hazak G 2016 {\it Phys. Rev. E} {\bf 94} 053210

\bibitem{Aberg2024} Aberg D, Grabowski P, Kruse, M and Wilson B G 2024 {\it High Energy Density Phys.} {\bf 50} 101079

\bibitem{Pain2023} Pain J-C 2023 {\it Opacity calculations: including more and more states}, APS Division of Plasma Physics Meeting Abstracts, NP11.137

\bibitem{Gilleron2004} Gilleron F and Pain J-C 2004 {\it Phys. Rev. E} {\bf 69} 056117

\bibitem{Pain2020} Pain J-C and Poirier M 2020 {\it J. Phys. B: At., Mol. Opt. Phys.} {\bf 53} 115002

\bibitem{Pain2025} Pain J-C, Aberg D and Wilson B G 2025 {\it High Energy Density Phys.} {\bf 54} 101174

\bibitem{Condon1959} Condon E U ans Shortley G H {\it The theory of atomic spectra} (Cambridge University Pres, London, 1959)

\bibitem{Sobelman1972}
Sobelman I I {\it Introduction to the Theory of Atomic Spectra} (Pergamon Press, Oxford, 1972)

\bibitem{Sobelman1992}
Sobelman I I {\it Atomic spectra and radiative transitions} (Springer-Verlag, Berlin, Heidelberg, 1992)

\bibitem{Zamick2005} Zamick L and Escuderos A 2005 {\it Phys. Rev. C} {\bf 71} 054308

\bibitem{Zamick2013} Zamick L and Escuderos A 2013 {\it Phys. Rev. C} {\bf 87} 044302 

\bibitem{Xin2022} Yin X and Zhao Y M 2022 {\it Chin. Phys. C} {\bf 46} 114101

\bibitem{Katriel1989} Katriel J and Novoselsky A 1989 {\it J. Phys. A: Math. Gen.} {\bf 22} 1245--1251

\bibitem{Zhao2005} Zhao Y M and Arima A 2005 {\it Phys. Rev. C} {\bf 72} 054307

\bibitem{Cleary1971} Cleary J G and Wybourne B G 1971 {\it J. Math. Phys.} {\bf 12} 45--52

\bibitem{Hirst1986} Hirst M G and Wybourne B G 1986 {\it J. Phys. A: Math. Gen.} {\bf 19} 1545--1549

\bibitem{Carlson1989} Carlson B V and Merchant A C 1989 {\it Phys. Rev. C} {\bf 40} 2265--2270

\bibitem{Subramanian1974} Subramanian P R and Devanathan V 1974 {\it J. Phys. A: Math. Nucl. Gen.} {\bf 16}, 1995--2007

\bibitem{Raychev1990} Raychev P P and Smirnov Y F 1990 {\it J. Phys. A: Math. Gen.} {\bf 23} 4417--4425

\bibitem{Planelles1996} Planelles J, Rajadell F, Karwowski J and Mas V 1996 {\it Phys. Rep.} {\bf 267}, 161--194

\bibitem{Pain2019} Pain J-C 2019 {\it Phys. Rev. C} {\bf 99} 054321

\bibitem{Poirier2021a} Poirier M and Pain J-C 2021 {\it J. Phys. B: At. Mol. Opt. Phys.} {\bf 54} 145002

\bibitem{Comtet1974} Comtet L {\it Advanced combinatorics; the art of finite and infinite expansions} (Dordrecht, Boston: D. Reidel Pub. Co., 1974).

\bibitem{Andrews1984} Andrews G A {\it The theory of partitions} (Cambridge University Press, 1984).

\bibitem{Stanley2011} Stanley R P {\it Enumerative combinatorics}, 2nd edition (Cambridge University Press, 2011).

\bibitem{Poirier2021b} Poirier M and Pain J-C 2021 {\it J. Phys. B: At. Mol. Opt. Phys.} {\bf 54} 145006

\bibitem{Wilf2005} Wilf H S {\it Generatingfunctionology}, 3rd Edition (A K Peters/CRC Press, 2005).

\bibitem{Poirier2021} Poirier M and Pain J-C 2021 {\it Phys. Rev. C} {\bf 104} 064324

\bibitem{Poirier2024} Poirier M and Pain J-C 2024 {\it Phys. Rev. C} {\bf 109} 024306

\bibitem{Sunko1985} Sunko D K and Svrtan D 1985 {\it Phys. Rev. C} {\bf 31} 1929--1933 

\bibitem{Sunko1986} Sunko D K 1986 {\it Phys. Rev. C} {\bf 33} 1811--1813

\bibitem{Gilleron2009} Gilleron F and Pain J-C 2009 {\it High Energy Density Phys.} {\bf 5} 320--327

\bibitem{Lain1998} Lain L and Torre A 1998 {\it J. Phys. B: At. Mol. Opt. Phys.} {\bf 31} 4259--4265

\bibitem{Gaussian} Quantum-chemistry software package {\it Gaussian}, \url{https://gaussian.com/}

\bibitem{ORCA} Quantum-chemistry software package {\it ORCA}, \url{https://www.faccts.de/orca/}

% Selection rules for the dematerialization of a particle into two photons
\bibitem{Yang1950} Yang C N 1950 {\it Phys. Rev.} {\bf 77} 242--245

% The moment of a 2-photon system
\bibitem{Landau1948} Landau L D 1948 {\it Dokl. Akad. Nauk SSSR} {\bf 60} 207--209

\bibitem{Zhao2003} Zhao Y M and Arima A 2003 {\it Phys. Rev. C} {\bf 68} 044310

\bibitem{Bao2016} Bao M, Zhao Y M and Arima A 2016 {\it Phys. Rev. C} {\bf 93} 014307

\bibitem{Faa} Fa\`a di Bruno F {\it Einleitung in die Theorie der Bin\"aren Formen} (B. G. Teubner, Leipzig, 1881), p. 126.

\end{thebibliography}
\end{document}